# ANTI-FERROELECTRIC THIN FILMS PHASE DIAGRAMS.

E.A.Eliseev*, M.D.Glinchuk, and A.N.Morozovska†

Institute for Problems of Materials Science, NAS of Ukraine,

Krjijanovskogo 3, 03142 Kiev, Ukraine

**Abstract**

In the paper we consider size effects of phase transitions and polar properties of the thin antiferroelectric films. We modified phenomenological approach proposed by Kittel. The Euler-Lagrange equations were solved by direct variational method. The free energy with renormalized coefficients depending on the film thickness has been derived. The approximate analytical expression for the coefficients dependence on film thickness, temperature, polarization gradient coefficient and extrapolation lengths were obtained. We have shown how the anti-ferroelectric “double” hysteresis loop transforms into the ferroelectric “single” one under the film thickness decrease. Proposed theoretical consideration explains the experimental results obtained in antiferroelectric $PbZrO_3$ thin films.

Keywords: antiferroelectrics, size effect, depolarization field, thin films.

PACS: 77.80.-e, 77.55.+f, 77.80.Bh,

## 1. Introduction

Antiferroelectrics are characterized by antiparallel aliment of local dipoles so that macroscopical polarization in the absence of electric field is zero. At the same time the polar state with parallel aliment have free energy close to the energy of antiparallel aliment, so that ferroelectric state can be reached under sufficiently high external electric field, mechanical stress etc. [1]. Antiferroelectric thin films and their heterostructures have numerous applications in microelectronic and microelectromechanical systems, capacitors, charge storage.

---

* Corresponding author e-mail: eliseev@i.com.ua

†. Permanent address: V. Lashkaryov Institute of Semiconductor Physics, NAS of Ukraine, 41, pr. Nauki, 03028 Kiev, Ukraine.

Mischenko et al. [2] have found giant electrocaloric effect in the thin films of $PbZr_{0.95}Ti_{0.05}O_3$. Chattopadhyay et al. [3] demonstrated that thin $PbZrO_3$ films on Si substrate possess switchable polarization. Hung et al. [4] have found that multilayer of $PbZrO_3/BaZrO_3$ has ferroelectric properties. Charnaya et al. [5] numerically calculated the phase diagrams for antiferroelectric confined systems of different shape in the absence of external electric field.

Ishchuk et al. [6] obtained that solid solutions $Pb_{1-3x/2}La_x(Zr_{1-y}Ti_y)O_3$ possess wide intervals of thermodynamic parameters in which the domains of the ferroelectric (FE) and antiferroelectric (AFE) phases coexist due to the small difference in their free energies.

Ayyub et al. [7] have shown that a typically antiferroelectric material $PbZrO_3$ or $BiNbO_4$ displays ferroelectric behavior below a critical film thickness characteristic of the system. Namely, 100nm $PbZrO_3$/Si film has ferroelectric hysteresis loop, whereas 900 nm film posses the antiferroelectric one. The antiferroelectric to ferroelectric transition takes place between 550nm and 400nm, i.e. it is typical size effect.

In the paper we consider size effects of phase transitions and polar properties of the thin antiferroelectric films and qualitatively explain size-induced the antiferroelectric-ferroelectric phase transition.

## 2. Free energy functional

Let us consider antiferroelectric thin film with the thickness $l$ $\left(-l/2 \leq z \leq l/2\right)$ on the thick substrate. In phenomenological theory approach free energy for this system can be written as the sum of bulk and surface parts $\Delta G = \Delta G_V + \Delta G_S$. For the description of the phase transitions in the antiferroelectric with two sublattices with polarizations $P_a$ and $P_b$ one can use the free energy expansion on $P_a, P_b$ powers similarly to the Kittel model [8]. However it is more convenient to switch to the new variables $P_A(z) = P_a(z) - P_b(z)$ and $P_F(z) = P_a(z) + P_b(z)$, which represent the "antipolar" and the "polar" order parameters respectively. For the case of polarization pointed perpendicular to the film surface, bulk part of the free energy functional with respect to the depolarization field and correlation energy acquires the following form:

$$\Delta G_V = \frac{1}{l}\int_{-l/2}^{l/2} dz \left[\begin{array}{l} \left(\frac{f}{2}+\frac{g}{4}\right)P_F^2(z)+\left(\frac{f}{2}-\frac{g}{4}\right)P_A^2(z)+\left(\frac{3h}{4}-\frac{q}{8}\right)P_F^2(z)\cdot P_A^2(z)+ \\ +\left(\frac{h}{8}+\frac{q}{16}\right)\left(P_F^4(z)+P_A^4(z)\right)-P_F(z)\left(E_0+\frac{1}{2}E_d\right)+ \\ +\frac{1}{2}\left(\delta+\frac{\xi}{2}\right)\left(\frac{dP_F(z)}{dz}\right)^2+\frac{1}{2}\left(\delta-\frac{\xi}{2}\right)\left(\frac{dP_A(z)}{dz}\right)^2 \end{array}\right] \quad (1)$$

Here $f, g, h, q$ are the coefficients before the free energy terms $P_{a,b}^2, P_aP_b, P_{a,b}^4, P_a^2P_b^2$ respectively; $\delta, \xi$ are the coefficients before the correlation energy terms $(\partial P_{a,b}/\partial z)^2, (\partial P_a/\partial z)(\partial P_b/\partial z)$ respectively; $E_0$ is the external electric field, $E_d$ is depolarization field. Hereinafter $h>0$, $q>-2h$, $\delta>0$, $|\xi|<2\delta$ and $g>0$ which corresponds to the antiferroelectric ordering in the bulk low temperature phase. Also we suppose $f$ as the linear function of temperature, namely $f-\frac{g}{2}=\mu(T-T_A)$. Here $\mu$ is related with the inverse Curie constant in paraelectric phase, $T_A$ is the temperature of transition from paraelectric to antiferroelectric phase for the bulk system.

The value $E_d$ for the case of single-domain insulator film covered with ideal electrodes can be written in the form [9]:

$$E_d = 4\pi\left(\overline{P}_F - P_F(z)\right). \quad (2)$$

Hereafter the bar over a physical quantity represents its spatial averaging over the film thickness.

Surface part of the free energy can be rewritten via $P_{A,F}^2$:

$$\Delta G_S = \frac{1}{2l}\left(\frac{\delta}{\lambda}+\frac{\xi}{2\zeta}\right)\left[P_F^2\left(\frac{l}{2}\right)+P_F^2\left(-\frac{l}{2}\right)\right]+\frac{1}{2l}\left(\frac{\delta}{\lambda}-\frac{\xi}{2\zeta}\right)\left[P_A^2\left(-\frac{l}{2}\right)+P_A^2\left(-\frac{l}{2}\right)\right] \quad (3)$$

The extrapolation lengths $\zeta>0$ and $\lambda>0$ can be only positive.

**3. Free energy with renormalized coefficients**

The coupled equations for the order parameters $P_F$ and $P_A$ can be obtained by variation over polarization of free energy functional (1) with respect to Eqs. (2), (3). This yields the following Euler-Lagrange equations with the boundary conditions:

$$\begin{cases} \left(f+\frac{g}{2}\right)P_F+\left(\frac{3h}{2}-\frac{q}{4}\right)P_A^2P_F+\left(\frac{h}{2}+\frac{q}{4}\right)P_F^3-\left(\delta+\frac{\xi}{2}\right)\frac{d^2P_F}{dz^2}=E_0+4\pi\left(\overline{P}_F-P_F\right), \\ \left.\left(\left(\frac{\delta}{\lambda}+\frac{\xi}{2\zeta}\right)P_F+\left(\delta+\frac{\xi}{2}\right)\frac{dP_F}{dz}\right)\right|_{z=l/2}=0, \quad \left.\left(\left(\frac{\delta}{\lambda}+\frac{\xi}{2\zeta}\right)P_F-\left(\delta+\frac{\xi}{2}\right)\frac{dP_F}{dz}\right)\right|_{z=-l/2}=0. \end{cases} \tag{4}$$

$$\begin{cases} \left(f-\frac{g}{2}\right)P_A+\left(\frac{3h}{2}-\frac{q}{4}\right)P_F^2P_A+\left(\frac{h}{2}+\frac{q}{4}\right)P_A^3-\left(\delta-\frac{\xi}{2}\right)\frac{d^2P_A}{dz^2}=E_A, \\ \left.\left(\left(\frac{\delta}{\lambda}-\frac{\xi}{2\zeta}\right)P_A+\left(\delta-\frac{\xi}{2}\right)\frac{dP_A}{dz}\right)\right|_{z=l/2}=0, \quad \left.\left(\left(\frac{\delta}{\lambda}-\frac{\xi}{2\zeta}\right)P_A-\left(\delta-\frac{\xi}{2}\right)\frac{dP_A}{dz}\right)\right|_{z=-l/2}=0. \end{cases} \tag{5}$$

Let us find as it was proposed earlier [10] the approximate solution of the nonlinear Eqs.(4), (5) by the direct variational method (see Appendix A). Briefly, averaging of the free energy (1-3) with appropriate trial functions (linearized solution of Eqs.(4), (5)) leads to the following form of the free energy with renormalized coefficients:

$$\Delta G(P_{VF},P_{VA})=\left[\frac{\alpha_F}{2}P_{VF}^2+\frac{\beta_F}{4}P_{VF}^4-P_{VF}E_0+\frac{\alpha_A}{2}P_{VA}^2+\frac{\beta_A}{4}P_{VA}^4+\frac{\eta}{2}P_{VF}^2P_{VA}^2\right]. \tag{6}$$

For the second order phase transitions the renormalized coefficients acquire the following form:

$$\alpha_F(T,l)\approx\mu T_A\left(\frac{T}{T_A}-1+\frac{g}{\mu T_A}+\frac{l_{cr}^F}{l}\right), \qquad l_{cr}^F=\frac{2\delta+\xi}{\mu T_A\lambda} \tag{7}$$

The coefficients before $P_{VA}^2$ has the form:

$$\alpha_A(T,l)\approx\mu T_A\left(\frac{T}{T_A}-1+\frac{{l_{cr}^A}^2}{l^2}\right), \qquad l_{cr}^A=\pi\sqrt{\left(\delta-\frac{\xi}{2}\right)\frac{1}{\mu T_A}}, \tag{8}$$

All other coefficients before $P_F$ powers have rather simple form only at $l/l_F>>1$:

$$\beta_F\approx\beta_A\approx\left(\frac{h}{2}+\frac{q}{4}\right), \qquad \eta\approx\left(\frac{3}{2}h-\frac{q}{4}\right). \tag{9}$$

The necessary conditions of the free energy (6) thermodynamic stability are the following:

$$\beta_F > 0, \quad \beta_A > 0, \quad \eta > -\sqrt{\beta_F \beta_A}\ . \tag{11}$$

The equations of state for the amplitudes $P_{VF,\ VA}$ can be obtained by variation of the renormalized free energy (8).

## 4. Phase diagrams and dielectric properties

### *4.1. Phase diagrams at zero external field*

Similarly to the bulk systems, at $E_0 = 0$ we obtained that film can be in different phases, namely: paraelectric PE-phase ($P_A = P_F = 0$), antiferroelectric AFE-phase ($P_A \neq 0,\quad P_F = 0$), ferroelectric FE-phase ($P_A = 0,\quad P_F \neq 0$), mixed ferroelectric FI-phase ($P_A \neq 0,\quad P_F \neq 0$) (see Appendix B for details).

In zero external field $E_0 = 0$ we found that two types of phase diagrams are possible, which are depicted in Figs.1(a, b).

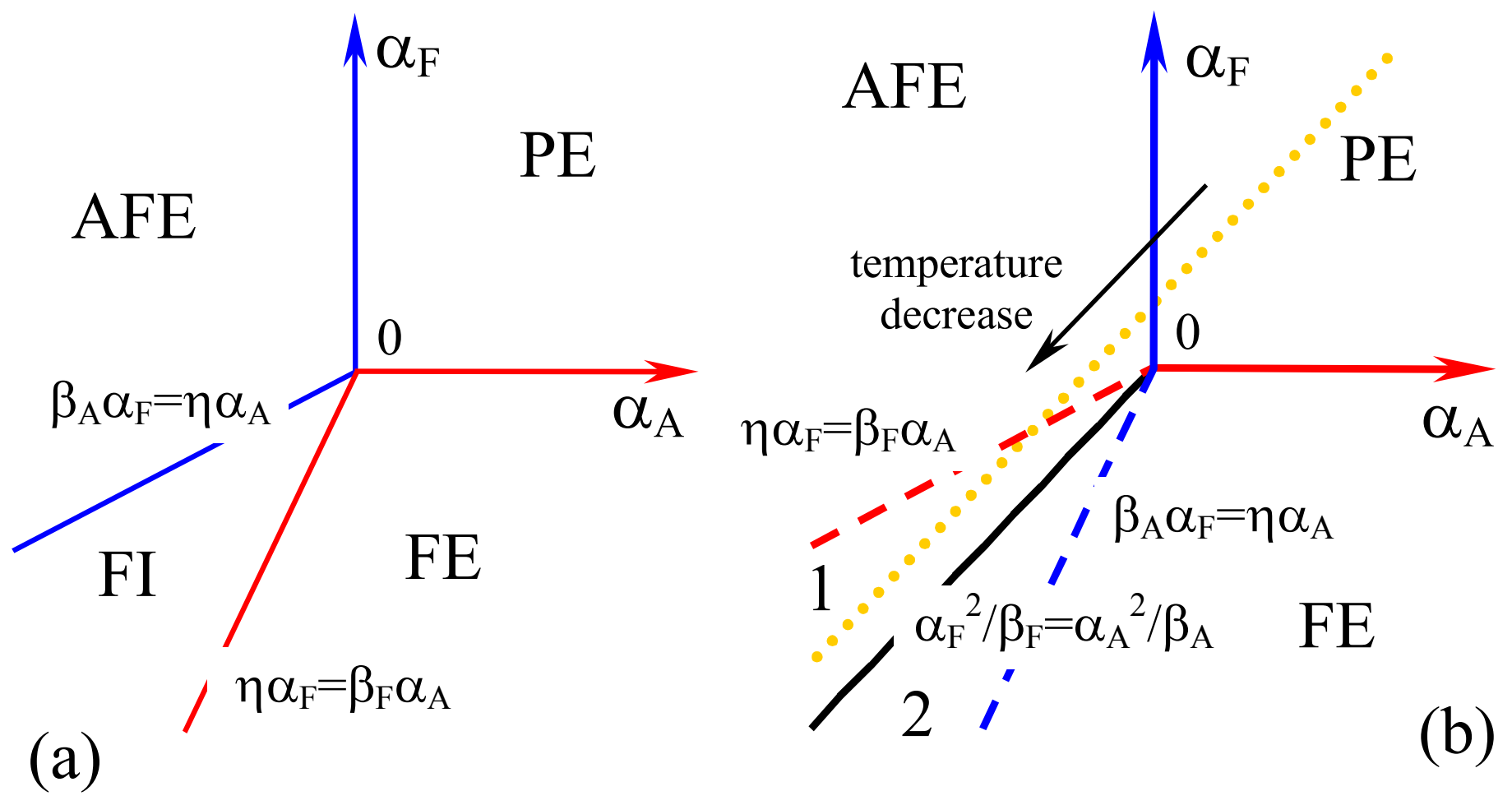

**FIG. 1.** Phase diagram in the coordinates $(\alpha_A, \alpha_F)$ for $\beta_A = \beta_F$, $\eta/\sqrt{\beta_A\beta_F} = 1/2$ (a) and $\eta/\sqrt{\beta_A\beta_F} = 2$ (b). Solid lines denote the phase transitions, dashed lines corresponds to the phases meta-stability limits if they do not coincide with the phase transitions points. Dotted line represents the temperature dependence of the free energy coefficients. In the regions 1 and 2 between solid and dashed curves in part (b) the absolute stable AFE (FE) phase coexists with metastable FE (AFE) phase respectively.

The boundary of PE phase is determined from the conditions $\alpha_F = 0$, $\alpha_A > 0$ or $\alpha_A = 0$, $\alpha_F > 0$. The boundary between FE and FI phases exists at $\beta_F\beta_A - \eta^2 > 0$ and is determined from the conditions $\alpha_F \le 0$ and $\alpha_A\beta_F = \alpha_F\eta$. The boundary between FI and AFE phases exists at $\beta_F\beta_A - \eta^2 > 0$ and is determined from the conditions $\alpha_A \le 0$ and $\alpha_F\beta_A = \alpha_A\eta$ (see Fig.1(a)). The boundary between FE and AFE phases exists at $\beta_F\beta_A - \eta^2 < 0$ and is determined from the conditions $\alpha_F \le 0$, $\alpha_A \le 0$, $\alpha_F^2/\beta_F = \alpha_A^2/\beta_A$ (see Fig. 1(b)).

Using the free energy coefficients (7), (9) dependences on the film thickness and temperature one can find that always $\alpha_A(T, l \to \infty) < \alpha_F(T, l \to \infty)$ as it should be expected for the bulk material, because AFE-phase is stable at $g > 0$. However, even rough estimation proves, that the different situation is possible for the thin films. For instance $\alpha_A(T,l) > 0$ and $\alpha_F(T,l) < 0$ under the conditions of small $g$ values and $\frac{2\delta + \xi}{\lambda} < l < \frac{\pi^2}{2}\lambda\frac{(2\delta - \xi)}{2\delta + \xi}$ ($T = 0$ K). The latter being possible at $\lambda > \frac{\pi(2\delta + \xi)}{\sqrt{2(2\delta - \xi)}}$. This means that FE-phase could be stable in the thin enough film under the following conditions on extrapolation length $\frac{\pi(2\delta + \xi)}{\sqrt{2(2\delta - \xi)}} < \lambda << \frac{\sqrt{\delta}}{\mu T_A}$. Simple estimation gives $\pi\sqrt{\delta} < \lambda << \frac{\sqrt{\delta}}{\mu T_A}$ or $3\,nm < \lambda << 100\ nm$. As an illustration, in Fig.2 we plotted the phase diagram in the coordinates temperature - film thickness. It is clear from Fig.2 that FE phase is absolutely stable at small thickness

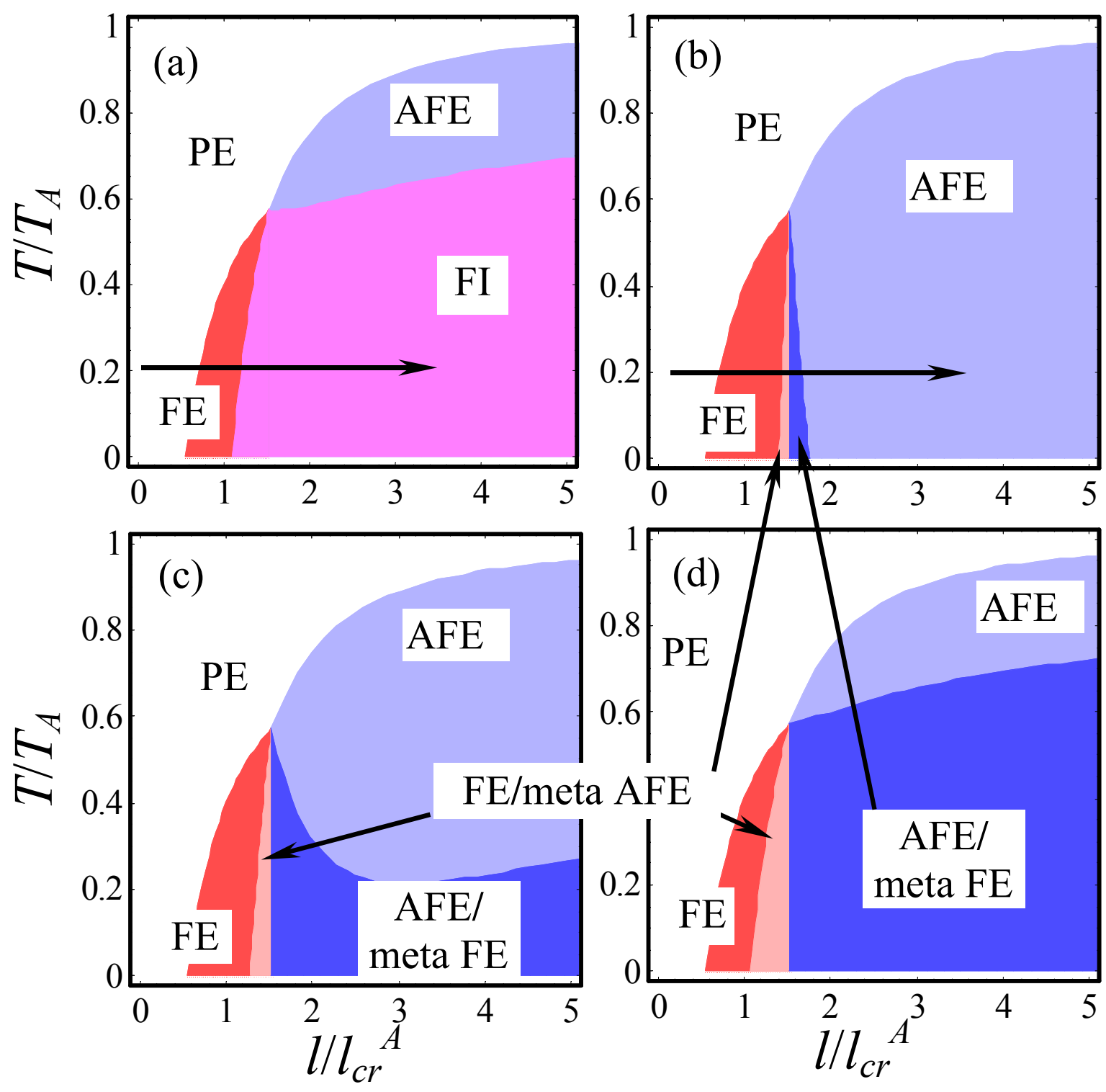

Fig.2. Phase diagram in the coordinates temperature-thickness for $l_{cr}^F / l_{cr}^A = 0.5$, $\beta_A = \beta_F$, $g/\mu T_A = 0.1$ and different ratio $\eta/\sqrt{\beta_A\beta_F} = 0.4, 1.1, 1.3, 3$ (a, b, c, d)

***4. 2 External field influence. Critical field of antiferroelectric-ferroelectric phase transition.***

The minimization of the free energy (8) at nonzero external filed $E_0 \neq 0$ allows one to obtain simple, but rather cumbersome equations for the dependence of order parameters on the external field. We plotted the hysteresis curves obtained in this way for different parameters values in Figs. 3-4. It is clear from Fig. 4 that the for the case $\eta/\sqrt{\beta_A\beta_F} > 1$ ferroelectric "single" hysteresis loop transforms into antiferroelectric "double" one the under the film thickness increase. While at $\eta/\sqrt{\beta_A\beta_F} \leq 1$ dependence of $P_F$ on external field so not reveal double hysteresis loops, only the field induced transition between mixed FI-phase and FE phase occurs. The former opportunity qualitatively explains the experimental results obtained in antiferroelectric $PbZrO_3$ thin films by Ayyub et al. [7] (see Fig.4)

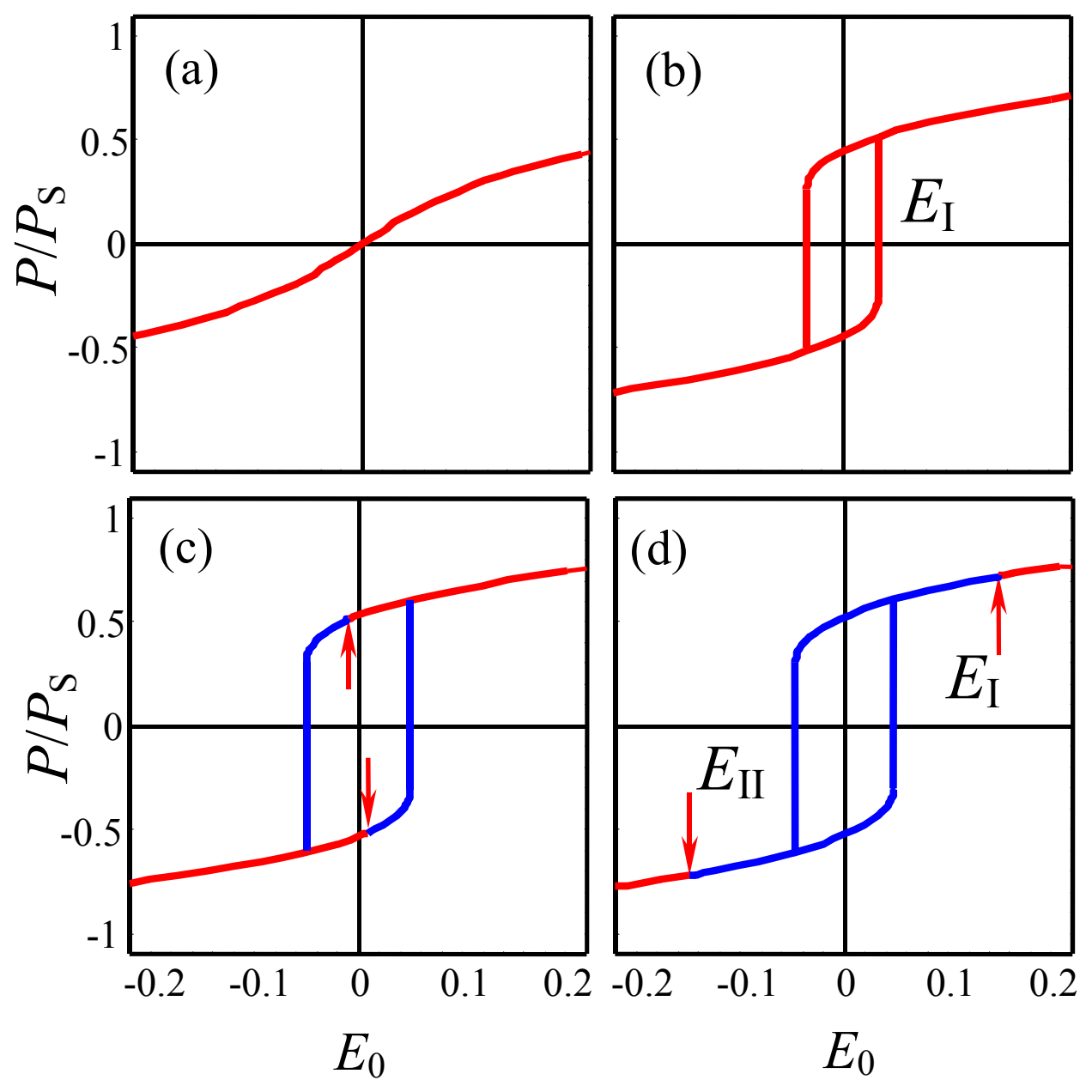

Fig. 3 Polarization dependence on external field for the following parameter values $l_{cr}^F/l_{cr}^A = 0.5$, $\eta/\sqrt{\beta_A\beta_F} = 0.4$, $T/T_A = 0.2$ and different film thickness $l/l_{cr}^A = 0.5, 1, 1.2, 1.3$.

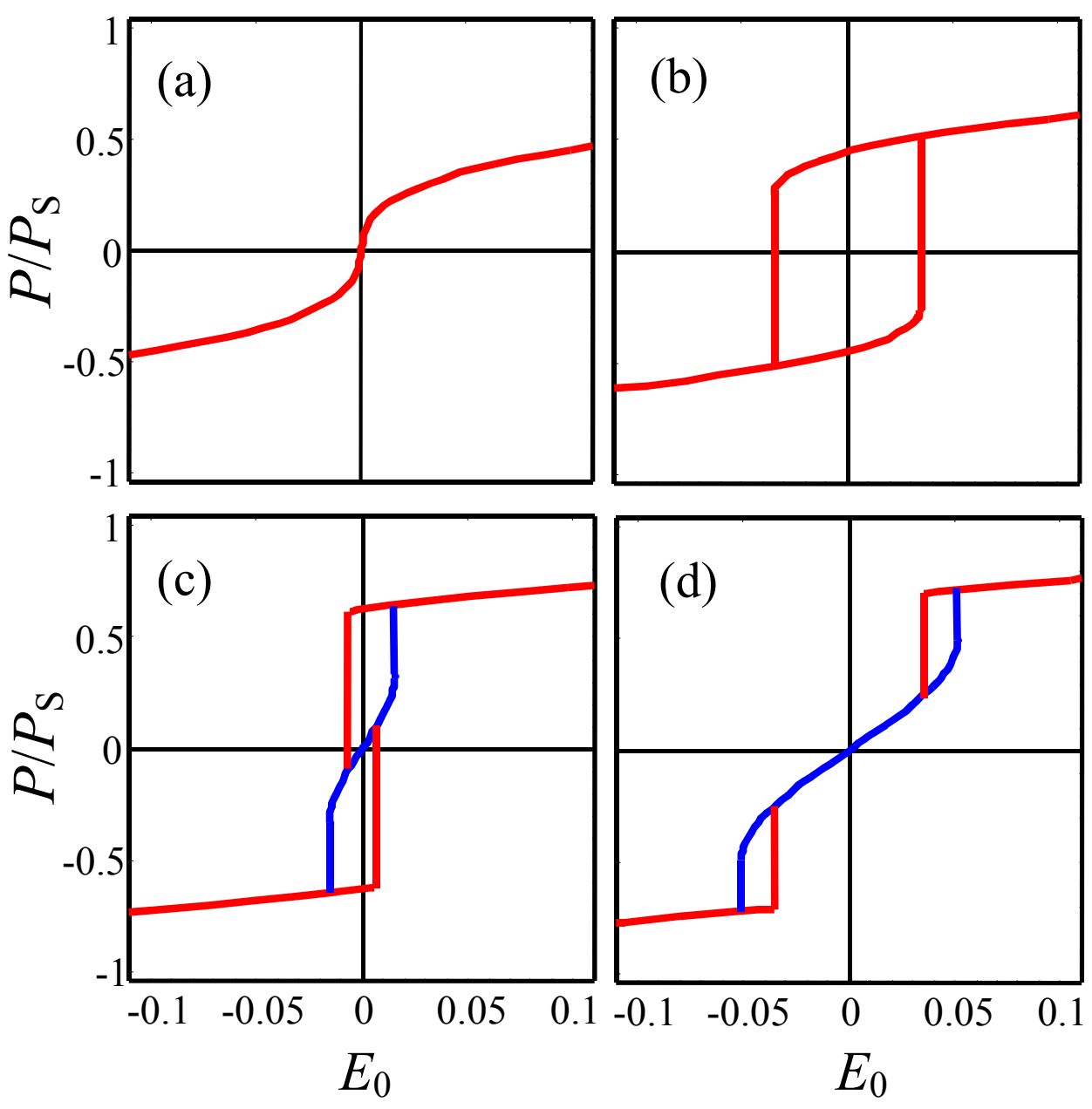

Fig. 4 Polarization dependence on external field for the following parameter values $l_{cr}^F/l_{cr}^A = 0.5$, $\eta/\sqrt{\beta_A\beta_F} = 1.1$, $T/T_A = 0.2$ and different film thickness $l/l_{cr}^A = 0.7, 1, 1.62, 2$.

In the case when bulk system does not reveal mixed FI phase at zero field ($\eta > \sqrt{\beta_F\beta_A}$) we obtained the critical fields that determine the antiferroelectric hysteresis loop position and width (see Figs. 3-4). Namely, the critical fields absolute values are

$$E_I = \sqrt{-\frac{\alpha_A}{\eta}}\left(\alpha_F - \alpha_A\frac{\beta_F}{\eta}\right), \quad E_{II} = \frac{2}{3}\sqrt{\frac{\beta_F\alpha_F - \eta\alpha_A}{3\left(\eta^2 - \beta_F\beta_A\right)}}\left(\alpha_F - \alpha_A\frac{\beta_F}{\eta}\right). \tag{12}$$

In the Fig. 5 the critical field dependence on the film thickness and temperature is presented. It is clear from Fig.5 that the critical fields reveal pronounced maximum caused by thickness dependences $1/l$ and $1/l^2$ in Eqs.(7) and (8) for $\alpha_F$ and $\alpha_A$ correspondingly.

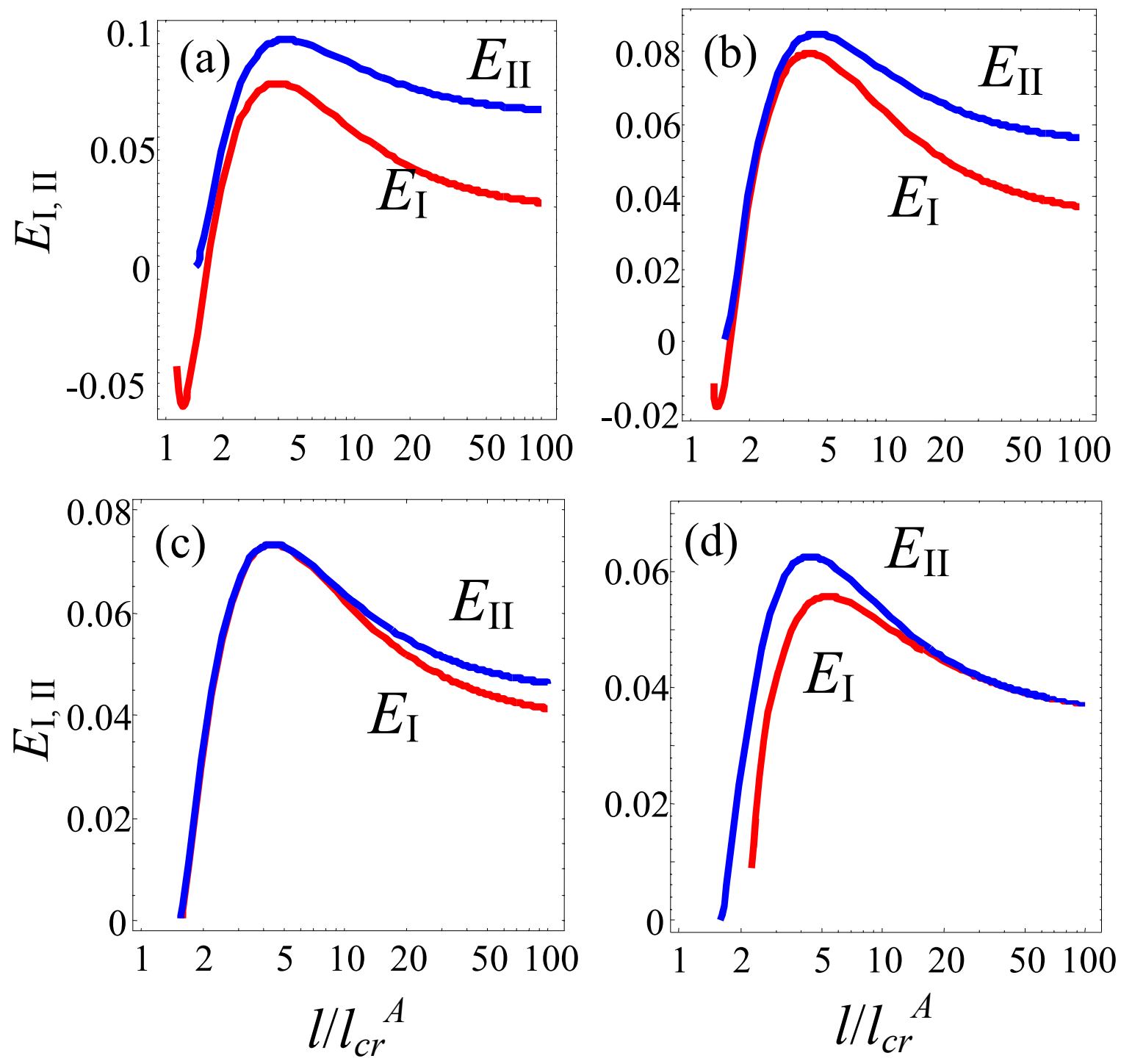

Fig. 5 Critical fields dependence on film thickness for $l_{cr}^F/l_{cr}^A = 0.5$, $\eta/\sqrt{\beta_A\beta_F} = 1.1$ and different temperatures $T/T_A = 0.2, 0.4, 0.6, 0.8$.

## 5. Conclusion

We modified approach proposed by Kittel for thermodynamical description of thin antiferroelectric films. The free energy with renormalized coefficients depending on the film thickness has been derived by direct variational method. The approximate analytical expression for the coefficients dependence on film thickness, temperature, polarization gradient coefficient and extrapolation lengths were obtained.

We have found conditions on the region of material parameters, under which the antiferroelectric “double” hysteresis loop transforms into the ferroelectric “single” one under the film thickness decrease.

Proposed theoretical consideration qualitatively explains the experimental results obtained in antiferroelectric $PbZrO_3$ thin films.

**Appendix A**

With respect to depolarization field and correlation effect the free energy $\Delta G = \Delta G_V + \Delta G_S$ expansion on sub-lattices polarizations powers has the view:

$$\Delta G_V = \frac{1}{l}\int_{-l/2}^{l/2} dz \begin{pmatrix} f\left(P_a^2 + P_b^2\right) + gP_a \cdot P_b + h\left(P_a^4 + P_b^4\right) + qP_a^2 \cdot P_b^2 \\ + \delta\left(\frac{dP_a}{dz}\right)^2 + \delta\left(\frac{dP_b}{dz}\right)^2 + \xi\left(\frac{dP_a}{dz}\right)\left(\frac{dP_b}{dz}\right) + \\ - P_a E_a - P_b E_b - 2\pi\left(P_a + P_b\right)\left(\overline{P}_a + \overline{P}_b - P_a - P_b\right) \end{pmatrix} \tag{A.1}$$

$$\begin{aligned} \Delta G_S = &\frac{\delta}{\lambda l}\left[P_a^2\left(\frac{l}{2}\right) + P_a^2\left(-\frac{l}{2}\right) + P_b^2\left(-\frac{l}{2}\right) + P_b^2\left(-\frac{l}{2}\right)\right] + \\ &+ \frac{\xi}{\zeta l}\left[P_a\left(\frac{l}{2}\right)P_b\left(\frac{l}{2}\right) + P_a\left(-\frac{l}{2}\right)P_b\left(-\frac{l}{2}\right)\right] \end{aligned} \tag{A.2}$$

For the description of the phase transitions in the antiferroelectric films with two lattices the free energy $\Delta G$ with respect to new variables $P_A(z) = P_a(z) - P_b(z)$ and $P_F(z) = P_a(z) + P_b(z)$ acquires the much simpler form, namely:

$$\Delta G_V = \frac{1}{l}\int_{-l/2}^{l/2} dz \begin{bmatrix} \left(\frac{f}{2} + \frac{g}{4}\right)P_F^2(z) + \left(\frac{f}{2} - \frac{g}{4}\right)P_A^2(z) + \left(\frac{3h}{4} - \frac{q}{8}\right)P_F^2(z)\cdot P_A^2(z) + \\ + \frac{1}{2}\left(\delta + \frac{\xi}{2}\right)\left(\frac{dP_F(z)}{dz}\right)^2 + \frac{1}{2}\left(\delta - \frac{\xi}{2}\right)\left(\frac{dP_A(z)}{dz}\right)^2 + \left(\frac{h}{8} + \frac{q}{16}\right)\left(P_F^4(z) + P_A^4(z)\right) + \\ - P_F(z)\left(E_0 + 2\pi\left(\overline{P}_F - P_F(z)\right)\right) - P_A(z)E_A \end{bmatrix} \tag{A.3}$$

$$\Delta G_S = \frac{1}{2l}\left(\frac{\delta}{\lambda} + \frac{\xi}{2\zeta}\right)\left[P_F^2\left(\frac{l}{2}\right) + P_F^2\left(-\frac{l}{2}\right)\right] + \frac{1}{2l}\left(\frac{\delta}{\lambda} - \frac{\xi}{2\zeta}\right)\left[P_A^2\left(-\frac{l}{2}\right) + P_A^2\left(-\frac{l}{2}\right)\right] \tag{A.4}$$

The coupled equations for the order parameters can be obtained by variation over polarization of free energy functional (A.3), (A.4). This yields the following Euler-Lagrange equations with the boundary conditions:

$$\begin{cases} \left(f + \frac{g}{2}\right)P_F + \left(\frac{3h}{2} - \frac{q}{4}\right)P_A^2 P_F + \left(\frac{h}{2} + \frac{q}{4}\right)P_F^3 - \left(\delta + \frac{\xi}{2}\right)\frac{d^2 P_F}{dz^2} = E_0 + 4\pi\left(\overline{P}_F - P_F\right), \\ \left.\left(\left(\frac{\delta}{\lambda} + \frac{\xi}{2\zeta}\right)P_F + \left(\delta + \frac{\xi}{2}\right)\frac{dP_F}{dz}\right)\right|_{z=l/2} = 0, \quad \left.\left(\left(\frac{\delta}{\lambda} + \frac{\xi}{2\zeta}\right)P_F - \left(\delta + \frac{\xi}{2}\right)\frac{dP_F}{dz}\right)\right|_{z=-l/2} = 0. \end{cases} \tag{A.5}$$

$$\begin{cases} \left(f-\dfrac{g}{2}\right)P_A+\left(\dfrac{3h}{2}-\dfrac{q}{4}\right)P_F^2P_A+\left(\dfrac{h}{2}+\dfrac{q}{4}\right)P_A^3-\left(\delta-\dfrac{\xi}{2}\right)\dfrac{d^2P_A}{dz^2}=E_A, \\ \left.\left(\left(\dfrac{\delta}{\lambda}-\dfrac{\xi}{2\zeta}\right)P_A+\left(\delta-\dfrac{\xi}{2}\right)\dfrac{dP_A}{dz}\right)\right|_{z=l/2}=0, \quad \left.\left(\left(\dfrac{\delta}{\lambda}-\dfrac{\xi}{2\zeta}\right)P_A-\left(\delta-\dfrac{\xi}{2}\right)\dfrac{dP_A}{dz}\right)\right|_{z=-l/2}=0. \end{cases} \tag{A.6}$$

We will choose the one-parametric trial functions similarly to [11], in the form of solutions of linearized Eqs.(A.5), (A.6) that satisfy the boundary conditions. Hereinafter we use the following trial functions:

$$P_F(z)=P_{VF}\frac{1-\varphi(z)}{1-\overline{\varphi(z)}} \qquad P_A(z)=P_{VA}\frac{1-\phi(z)}{1-\overline{\phi(z)}}. \tag{A.7}$$

The variational parameters - amplitudes $P_{VF,A}$ must be determined by the minimization of the free energy (A.3). Hereinafter we used the following functions:

$$\varphi(z)=\frac{ch(z/l_F)}{ch(l/2l_F)+(\lambda_F/l_F)sh(l/2l_F)},$$
$$\lambda_F=\left(\delta+\frac{\xi}{2}\right)\Big/\left(\frac{\delta}{\lambda}+\frac{\xi}{2\zeta}\right), \qquad l_F=\sqrt{\left(\delta+\frac{\xi}{2}\right)\Big/\left(4\pi+f+\frac{g}{2}\right)}\approx\sqrt{\frac{\delta}{4\pi}+\frac{\xi}{8\pi}}. \tag{A.8}$$

$$\phi(z)=\begin{cases} \dfrac{ch(z/l_A)}{ch(l/2l_A)+(\lambda_A/l_A)sh(l/2l_A)}, & f-\dfrac{g}{2}>0 \\ \dfrac{\cos(z/l_A)}{\cos(l/2l_A)-(\lambda_A/l_A)\sin(l/2l_A)}, & f-\dfrac{g}{2}<0 \end{cases}$$
$$\lambda_A=\left(\delta-\frac{\xi}{2}\right)\Big/\left(\frac{\delta}{\lambda}-\frac{\xi}{2\zeta}\right), \qquad l_A=\sqrt{\left(\delta-\frac{\xi}{2}\right)\Big/\left|f-\frac{g}{2}\right|}. \tag{A.9}$$

Here the ferroelectric and antiferroelectric extrapolation ($\lambda_F$, $\lambda_A$) and characteristic ($l_F$, $l_A$) lengths are introduced: For the majority of ferroelectrics $l_F\sim 1\div 10\,\mathring{A}$, while the situation with $l_A$ is more complex. In the case $(f-g/2)<0$ Eq.(A.9) could be used as trial function only when its denominator is a positive finite quantity. Thus, the critical thickness $l_{cr}^A(T)$ exists. It is the minimum root of equation $tg\left(l_{cr}^A/2l_A(T)\right)=l_A(T)/\lambda_A$.

The renormalized coefficients acquire the following form:

$$\alpha_F(T,l) = \frac{(f+g/2)+4\pi\overline{\varphi(z)}}{1-\overline{\varphi(z)}} \approx \left(f+\frac{g}{2}\right)+4\pi\overline{\varphi} \qquad \text{(A.10)}$$

$$\overline{\varphi(l)} = \frac{2l_F}{l}\cdot\frac{sh(l/2l_F)}{ch(l/2l_F)+(\lambda_F/l_F)sh(l/2l_F)} \approx \begin{cases} \dfrac{2l_F^2}{l(l_F+\lambda_F)}, & l/l_F >> 1 \\ \dfrac{l_F}{l_F+\lambda_F(l/2l_F)}, & l/l_F << 1 \end{cases} \qquad \text{(A.11)}$$

Where $\lambda_F = \left(\delta+\frac{\xi}{2}\right)\Big/\left(\frac{\delta}{\lambda}+\frac{\xi}{2\zeta}\right)$, $l_F = \sqrt{\left(\delta+\frac{\xi}{2}\right)\Big/\left(4\pi+f+\frac{g}{2}\right)} \approx \sqrt{\frac{\delta}{4\pi}+\frac{\xi}{8\pi}}$. It is seen that always $0<\overline{\varphi}<1$ at $\lambda_F>0$, therefore introduced in the denominator of Eq.(A.10) multiplier $(1-\overline{\varphi})$ is positive and could not be zero, thus it does not change $\alpha_F$ sign.

The coefficients before $P_{VA}^2$ obtained with the help of trial function (A.9) have rather simple form only at $(\lambda_A/l_A)^2 >> 1$ or $(\lambda_A/l_A)^2 << 1$, namely the [1/1]-Pade approximation over $1/l$ powers has the form:

$$\alpha_A(T,l) = \frac{f-\frac{g}{2}}{1-\overline{\phi(z)}} \approx \begin{cases} \left(f-\frac{g}{2}\right)\left(1-\frac{2l_A}{\lambda_A}\frac{l_A}{l}\right) = f-\frac{g}{2}+\left(\frac{\delta}{\lambda}-\frac{\xi}{2\zeta}\right)\frac{2}{l}, & \left(\frac{\lambda_A}{l_A}\right)^2 >> 1 \\ \left(f-\frac{g}{2}\right)\left(1-\left(\frac{\pi l_A}{l}\right)^2\right) = f-\frac{g}{2}+\left(\delta-\frac{\xi}{2}\right)\frac{\pi^2}{l^2}, & \left(\frac{\lambda_A}{l_A}\right)^2 << 1 \end{cases}, \qquad \text{(A.12)}$$

Where $\lambda_A = \left(\delta-\frac{\xi}{2}\right)\Big/\left(\frac{\delta}{\lambda}-\frac{\xi}{2\zeta}\right)$, $l_A = \sqrt{\left(\delta-\frac{\xi}{2}\right)\Big/\left|f-\frac{g}{2}\right|}$.

The free energy coefficients (A.11) (A.12) dependences on the film thickness and temperature are following:

$$\begin{aligned} \alpha_F(T,l) &= \mu(T-T_A)+g+\frac{8\pi l_F^2}{(l_F+\lambda_F)l} \approx \mu T_A\left(\frac{T}{T_A}-1+\frac{g}{\mu T_A}+\frac{1}{\mu T_A}\frac{2\delta+\xi}{\lambda l}\right) = \\ &= \mu T_A\left(\frac{T}{T_A}-1+\frac{g}{\mu T_A}+\frac{l_{cr}^F}{l}\right), \quad l_{cr}^F = \frac{2\delta+\xi}{\mu T_A \lambda} \end{aligned} \qquad \text{(A.13)}$$

$$\alpha_A(T,l) \approx \mu(T-T_A) + \left(\delta - \frac{\xi}{2}\right)\frac{\pi^2}{l^2} = \mu T_A\left(\frac{T}{T_A} - 1 + \frac{\pi^2}{2\mu T_A}\cdot\frac{2\delta-\xi}{l^2}\right) =$$
$$= \mu T_A\left(\frac{T}{T_A} - 1 + \frac{{l_{cr}^A}^2}{l^2}\right), \quad l_{cr}^A = \pi\sqrt{\left(\delta - \frac{\xi}{2}\right)\frac{1}{\mu T_A}} \qquad \text{(A.14)}$$

Expression (A.14) is valid at $\mu T_A \lambda << \sqrt{\delta}$ and $\lambda_F \approx \lambda$ that is true for most of ferroelectric materials.

**Appendix B**

Using the conditions of positively defined matrix of the free energy second derivatives, we obtained the following stability limits for the different phases. a). Paraelectric PE-phase ($P_A = P_F = 0$). The conditions of PE-phase *stability* are the following

$$\alpha_F \geq 0 \text{ and } \alpha_A \geq 0 \qquad \text{(B.1)}$$

b). Antiferroelectric AFE-phase ($P_A \neq 0, \quad P_F = 0$) exists at $\alpha_A < 0$. The conditions of AFE-phase *meta- or absolute stability* are the following

$$\alpha_A < 0 \text{ and } \alpha_F\beta_A - \alpha_A\eta > 0. \qquad \text{(B.2)}$$

c). Ferroelectric FE-phase ($P_A = 0, \quad P_F \neq 0$) exists at $\alpha_F < 0$. The conditions of FE-phase *meta- or absolute stability* ($E_0 = 0$) are the following

$$\alpha_F < 0 \text{ and } \alpha_A\beta_F - \alpha_F\eta > 0. \qquad \text{(B.3)}$$

d). Mixed ferroelectric FI-phase ($P_A \neq 0, \quad P_F \neq 0$). The conditions of FI-phase *meta- or absolute stability* ($E_0 = 0$) are the following

$$\beta_F\beta_A - \eta^2 > 0, \qquad \alpha_F\beta_A - \alpha_A\eta < 0, \qquad \alpha_A\beta_F - \alpha_F\eta < 0 \qquad \text{(B.4)}$$

The boundary of PE phase is determined from the conditions $\alpha_F = 0$, $\alpha_A > 0$ or $\alpha_A = 0$, $\alpha_F > 0$.

The boundary between FE and FI phases exists at $\beta_F\beta_A - \eta^2 > 0$ and is determined from the conditions $\alpha_F \leq 0$ and $\alpha_A\beta_F = \alpha_F\eta$.

The boundary between FI and AFE phases exists at $\beta_F\beta_A - \eta^2 > 0$ and is determined from the conditions $\alpha_A \leq 0$ and $\alpha_F\beta_A = \alpha_A\eta$.

The boundary between FE and AFE phases exists at $\beta_F\beta_A-\eta^2<0$ and is determined from the conditions $\alpha_F\le 0$, $\alpha_A\le 0$, $\alpha_F^2/\beta_F=\alpha_A^2/\beta_A$.

Using the necessary conditions of the phase stability, namely $P_{VF}^2>-\frac{\alpha_F}{3\beta_F}$, $P_{VF}^2>-\frac{\alpha_A}{\eta}$, $\alpha_F<0$ (for FE phase) and $P_{VF}^2<\frac{\alpha_F\beta_A-\alpha_A\eta}{3\left(\eta^2-\beta_F\beta_A\right)}$, $P_{VF}^2<-\frac{\alpha_A}{\eta}$, $\alpha_A<0$ (for FI phase) in the case when bulk system do not reveal mixed FI phase at zero field ($\eta>\sqrt{\beta_F\beta_A}$) we obtained the critical fields that determine the antiferroelectric hysteresis loop position and width. Namely, the critical fields absolute values are

$$E_I=\sqrt{-\frac{\alpha_A}{\eta}}\left(\alpha_F-\alpha_A\frac{\beta_F}{\eta}\right),\quad E_{II}=\frac{2}{3}\sqrt{\frac{\beta_F\alpha_F-\eta\alpha_A}{3\left(\eta^2-\beta_F\beta_A\right)}}\left(\alpha_F-\alpha_A\frac{\beta_F}{\eta}\right). \tag{B.5}$$

**References**

---